\documentclass[pss]{wiley2sp} 
\pdfoutput=1
\usepackage[latin1]{inputenc}
\usepackage{amsmath}
\newcommand{\ph}[1]{\,\ensuremath{\mathrm{#1}}}
\DeclareGraphicsExtensions{.pdf}

\begin{document}
	\hyphenation{li-tho-gra-phy}

\title{Comprehensive characterization of an individual carbon nanotube transport device}

\titlerunning{Characterizing individual CNTs}

\author{%
  Robert Frielinghaus\textsuperscript{\textsf{\bfseries 1}},
  Karin Goß\textsuperscript{\textsf{\bfseries 1}},
  Stefan Trellenkamp\textsuperscript{\textsf{\bfseries 2}}, 
  Lothar Houben\textsuperscript{\textsf{\bfseries 3}},
  Claus M. Schneider\textsuperscript{\textsf{\bfseries 1}},
  Carola Meyer\textsuperscript{\Ast,\textsf{\bfseries 1}}
}
\authorrunning{Robert Frielinghaus et al.}

\mail{e-mail
  \textsf{c.meyer@fz-juelich.de}}

\institute{\flushleft%
  \textsuperscript{1}\,Peter Grünberg Institut (PGI-6) and JARA Jülich-Aachen Research Alliance, Forschungszentrum Jülich, 52425 Jülich, Germany\\
  \textsuperscript{2}\,Peter Grünberg Institut (PGI-8-PT) and JARA Jülich-Aachen Research Alliance, Forschungszentrum Jülich, 52425 Jülich, Germany\\
  \textsuperscript{3}\,Peter Grünberg Institut (PGI-5), Ernst Ruska Center for Microscopy and Spectroscopy with Electrons and JARA Jülich-Aachen Research Alliance, Forschungszentrum Jülich, 52425 Jülich, Germany\\}



\abstract{%
%
%
%
\abstcol{%
	We present a comprehensive characterization of an individual multiwalled carbon nanotube transport device combining electron microscopy and Raman spectroscopy with electrical measurements. Each method gives complementary information that mutually help to interpret each other.
  }{
  A sample design that allows for combining these investigation methods on individual carbon nanotube devices is introduced. This offers a direct correlation of transport features and shifts of Raman modes with structural properties as e.g. the contact interface and the morphology of the nanotube.
	}}



\maketitle		

\section{Introduction}
	Carbon nanotubes (CNTs) can be chemically functionalized in many ways by attaching molecules to the outside wall, to the tube ends or inside the inner hollow~\cite{Spudat2009}. Such modification of CNT transport devices leads to new applications, as e.g. chemical sensors or detectors~\cite{Qi2003}. In order to understand the changes in transport behavior due to chemical functionalization, information about the atomic structure of the individual device is required. 
	
	Here we present an approach to perform transport measurements, transmission electron microscopy (TEM) as well as Raman spectroscopy all on a single CNT device. These techniques provide complementary information as e.g. the structural parameters measured with TEM and the vibrational couplings measured with Raman spectroscopy~\cite{Spudat2010}. Transport measurements probe inter alia the electronic states of the CNT system and depend strongly on the degree of chemical functionalization~\cite{Eliasen2010} and the molecular environment~\cite{Goss2010}. 

\section{Experimental details}
	As first step in the preparation protocol electron beam lithography patterning and reactive ion beam etching (RIBE) open windows on a Si$_3$N$_4$ TEM membrane. CNTs are then grown by means of chemical vapor deposition (CVD) on well-defined places and are finally contacted. With this method we obtain isolated (functionalized) CNTs that are ready for correlated transport, Raman and TEM measurements. This process is not limited to CNTs, but may readily be applied to other nanomaterial systems, single molecular devices, nanowires etc.
	\subsection{Membrane patterning}\label{patterning}
		\begin{figure*}[htb]%
			\subfloat[Hole pattern with marker structure and observation slits.]{\label{Ubersicht}
				\includegraphics*[bb=-0 0 577 432,width=.3\textwidth]{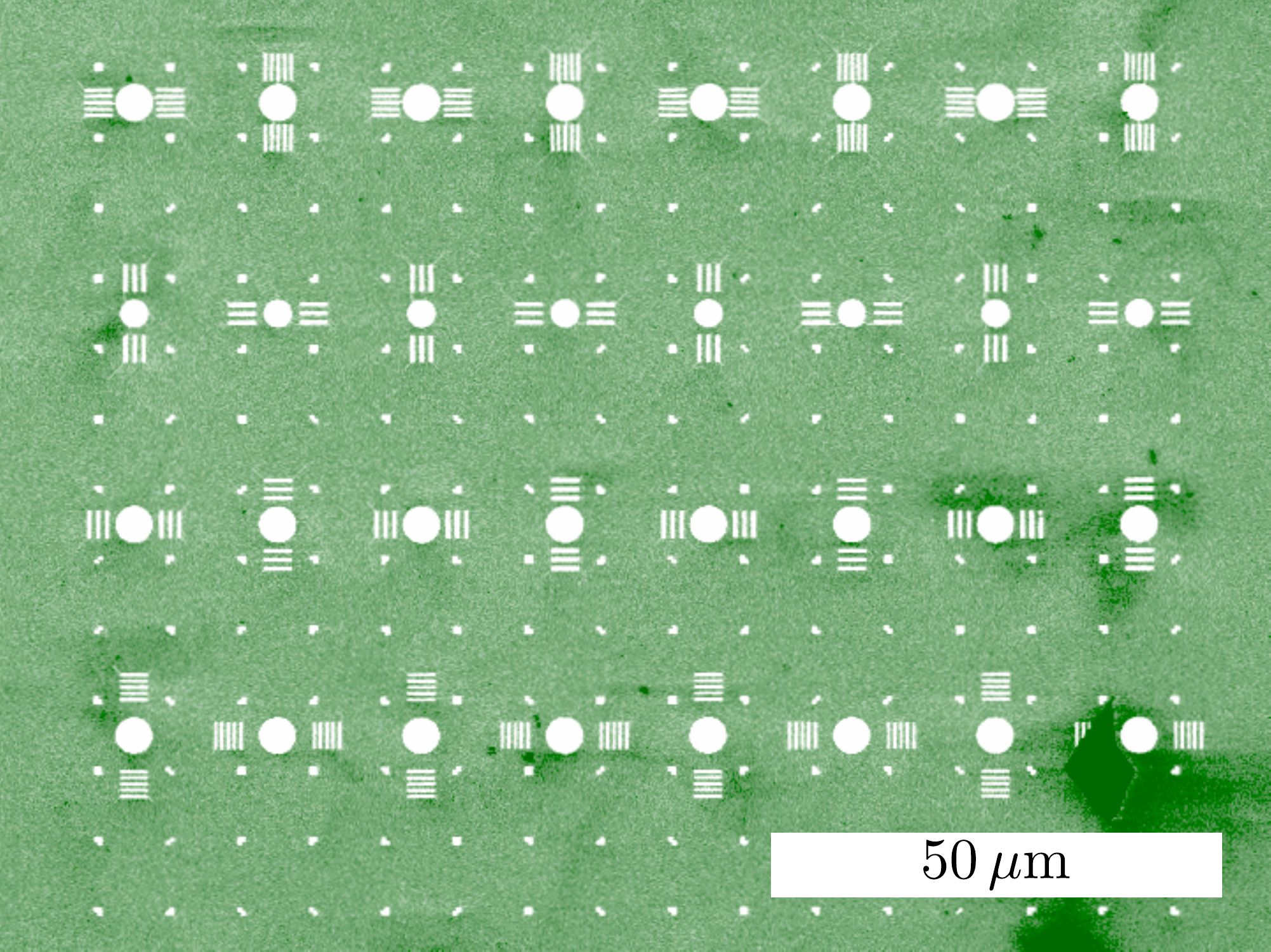}}\hfill
			\subfloat[As-grown CNTs cross the stripes (cf. arrows). Markers are needed to identify the field.]{\label{CNTverlauf}
				\includegraphics*[bb=0 0 1057 791=,width=.3\textwidth]{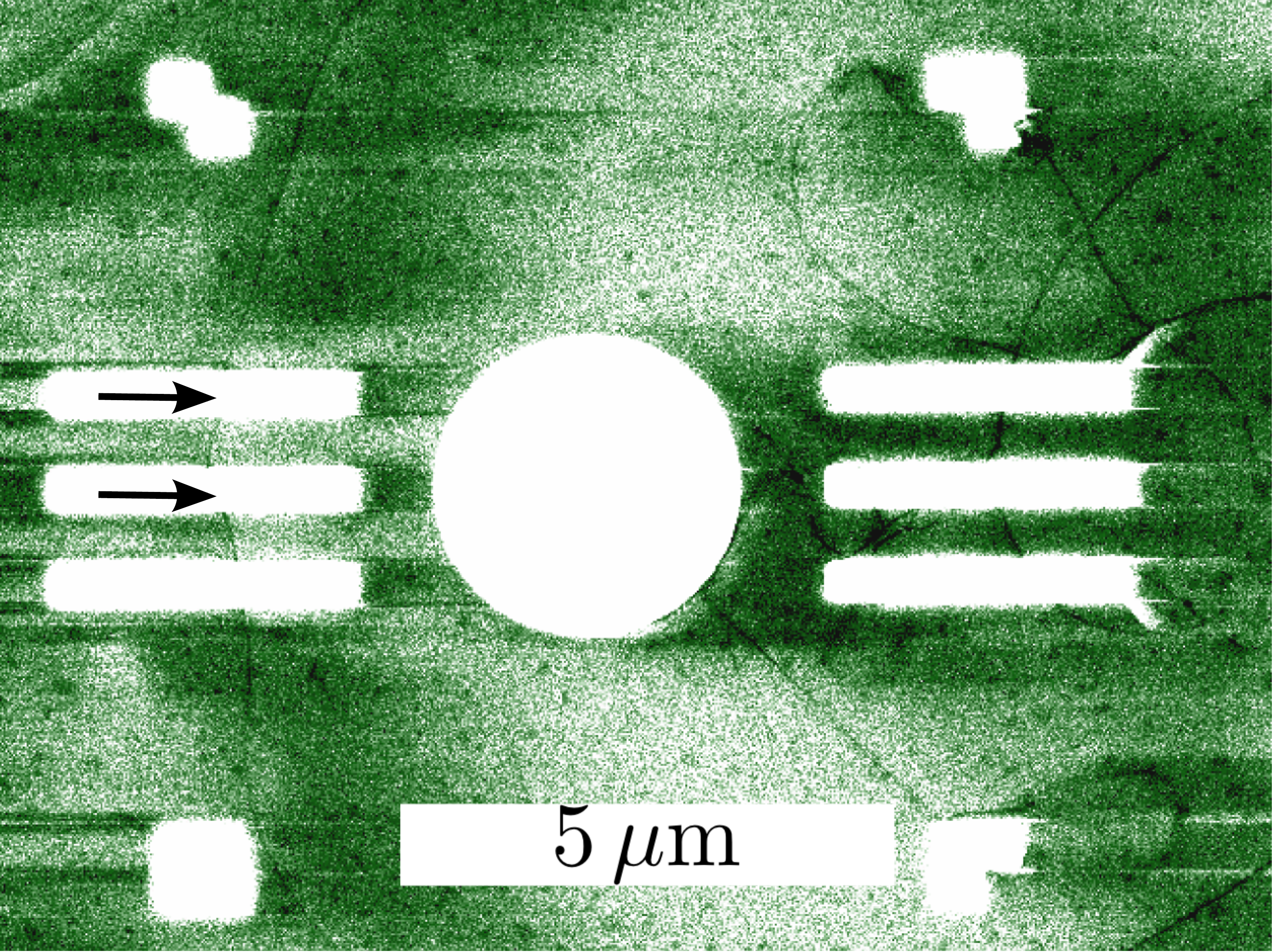}}\hfill
			\subfloat[Ti/Au contacts on the CNT marked with arrows corresponding to Fig.~\ref{CNTverlauf}. Inset: Schematics of the device]{\label{Kontaktiert}
				\includegraphics*[bb=0 0 1137 851,width=.3\textwidth]{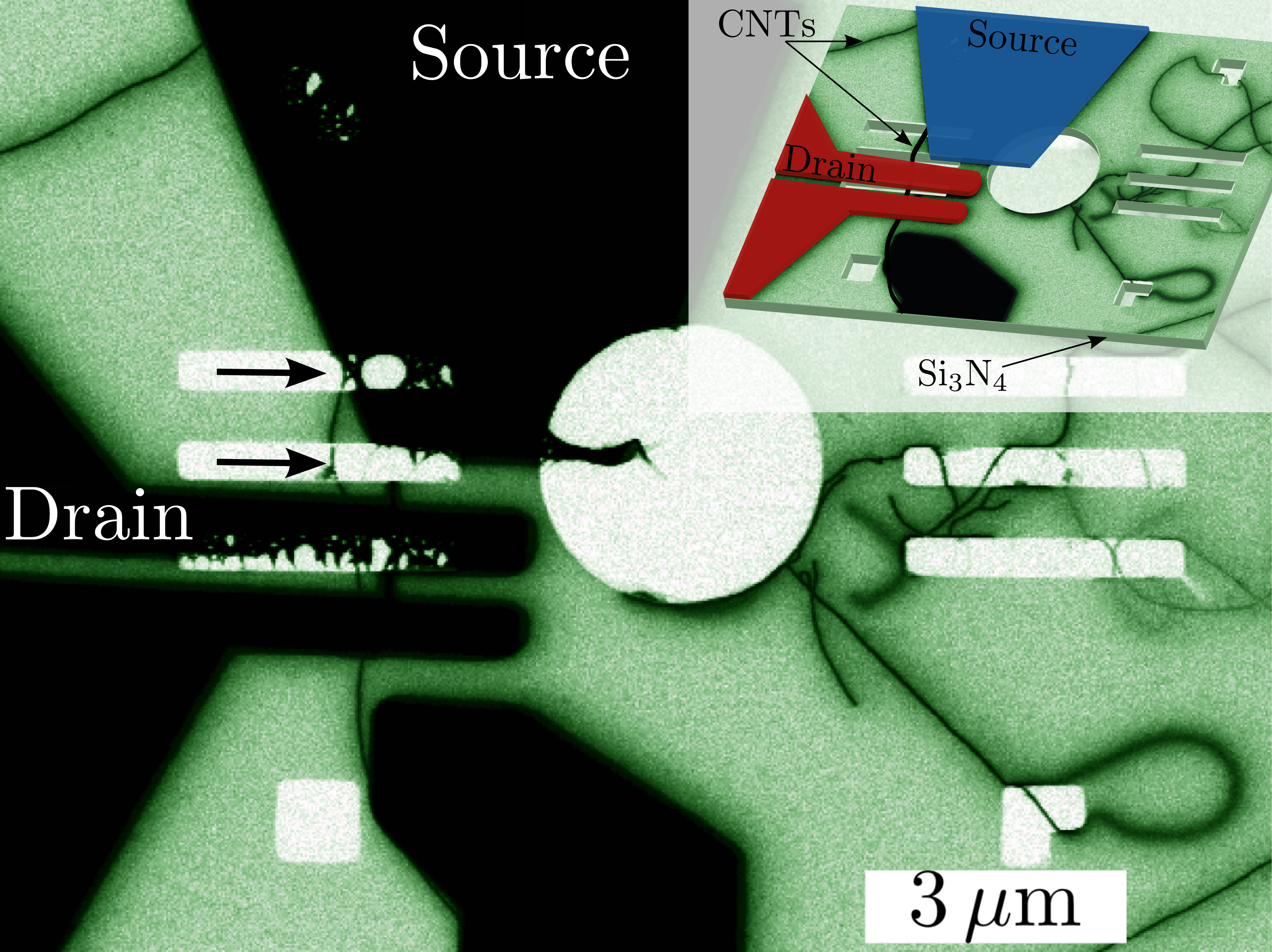}}%
			\caption{
				SEM images of the fabrication steps of the TEM membrane taken with 5\ph{kV} acceleration voltage. To enhance the contrast, regions with large secondary electron emission appear in black, those with no secondary electrons are white.}
		\end{figure*}
		The underlying sample is a DuraSiN\texttrademark~DTF-2523 TEM membrane. It consists of a round Si frame of 2.65\ph{mm} diameter and 300\ph{\mu m} thickness with a square central opening of $500\ph{\mu m}\times 500\ph{\mu m}$ that is covered with a 200\ph{nm} thick Si$_3$N$_4$ membrane. Before spincoating the sample needs to be treated with an O$_2$ plasma (300\ph{W}, 2\ph{min}) to enhance resist adhesion. We use a 600\ph{K} PMMA electron beam resist with 7\% solids content from AllResist (AR-P~669.07) that we spincoat at 7,000\ph{min^{-1}} for 35\ph{s}.
		
		An observation hole and marker pattern as shown in Fig.~\ref{Ubersicht} is written  on the membrane via electron beam lithography. The stripes have a nominal width of 200-400\ph{nm} and are suitable for TEM observation since the short suspended region prevents strong vibrations of the CNT. Furthermore, quantum dots can be formed by metal contacts on the tube with a side gate close enough to tune the electrochemical potential on the dot~\cite{Langer1996}. The round holes have a diameter of 1-2\ph{\mu m} and can be used for confocal Raman imaging. The markers are needed for the alignment of the contacts. The developed sample is exposed to a SF$_6$ RIBE for 6:30\ph{min} which selectively etches the Si$_3$N$_4$ of the membrane. Although the sample is placed in acetone overnight and treated with an O$_2$ plasma (600\ph{W}, 2\ph{h}) the etched PMMA cannot fully be removed. 

	\subsection{CNT growth and contacting}
		Relative to the hole pattern, some catalyst islands are deposited onto the sample using the same electron beam lithography process described above. We use an Fe/Mo catalyst and a growth temperature of 860\ph{^\circ C}~\cite{Spudat2009}. The CNTs are located using a scanning electron microscope (SEM) under ultra high vacuum conditions to avoid contamination with amorphous carbon. The resulting device is shown in Fig.~\ref{CNTverlauf}.
		
		In a third electron beam lithography step, contacts are patterned onto the sample (Fig.~\ref{Kontaktiert}). We use a 5\ph{nm}/60\ph{nm} Ti/Au bilayer that is known to form quantum dots in CNTs at low temperatures~\cite{Goss2010}. 
	\subsection{Raman and TEM measurements}
		Confocal Raman imaging is performed with an Ar ion laser operated at a wavelength $\lambda_0=488\ph{nm}$ and a Jobin Yvon T64000 spectrometer. The laser spot has a size of $\sim 2\ph{\mu m}$ and a power $P=420\ph{\mu W}$ to avoid nanotube destruction.

		High-Resolution Transmission Electron Microscopy measurements (HR-TEM) are performed in a FEI Titan 80-300 microscope equipped with a double-hexapole aberration corrector~\cite{Haider1998}. The acceleration voltage is chosen as 80 kV to reduce damages induced by the electron irradiation. The microscope is operated with an overfocus to ensure bright atom contrast. The TEM micrographs are recorded as a last measurement step, because it may be destructive albeit the low acceleration voltage~\cite{Meyer2009}. 
\section{Results}
	\begin{figure*}[htb]%
		\subfloat[Detailed SEM image of the measured device. The arrows indicate structures bridging the slit and only faintly visible in SEM; the blue rectangle the part analyzed in Fig.~\ref{TEMscan}]{\label{REMpos30}
			\includegraphics*[bb=0 0 473 473,width=.32\textwidth]{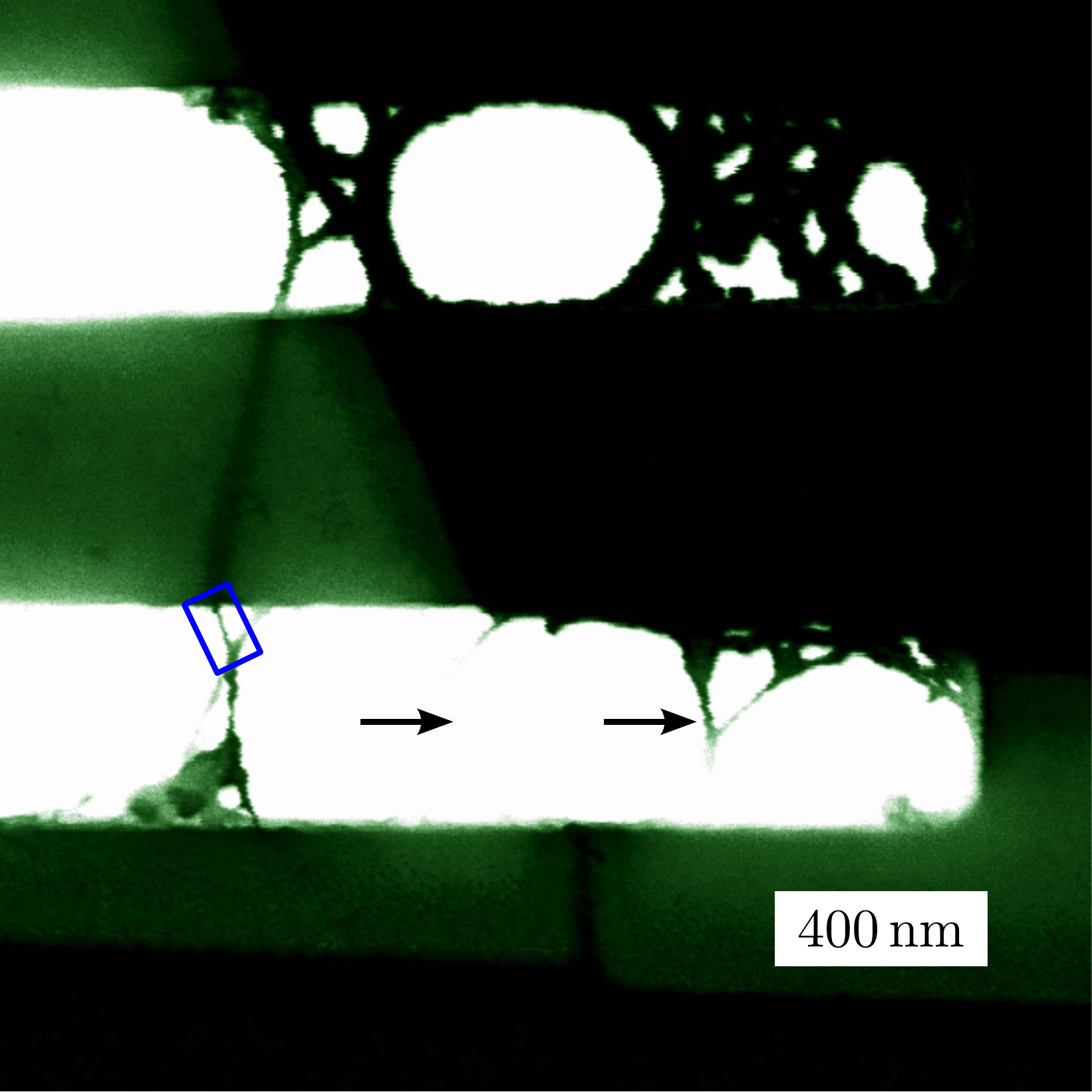}}\hfill
		\subfloat[Corresponding HR-TEM image to Fig.~\ref{REMpos30}. The dashed line indicates the CNT position before destruction by the TEM electron beam. Arrows point to the same structures as in Fig.~\ref{REMpos30}]{\label{TEMgesamt}
			\includegraphics*[bb=0 0 1354 1337,width=.32\textwidth]{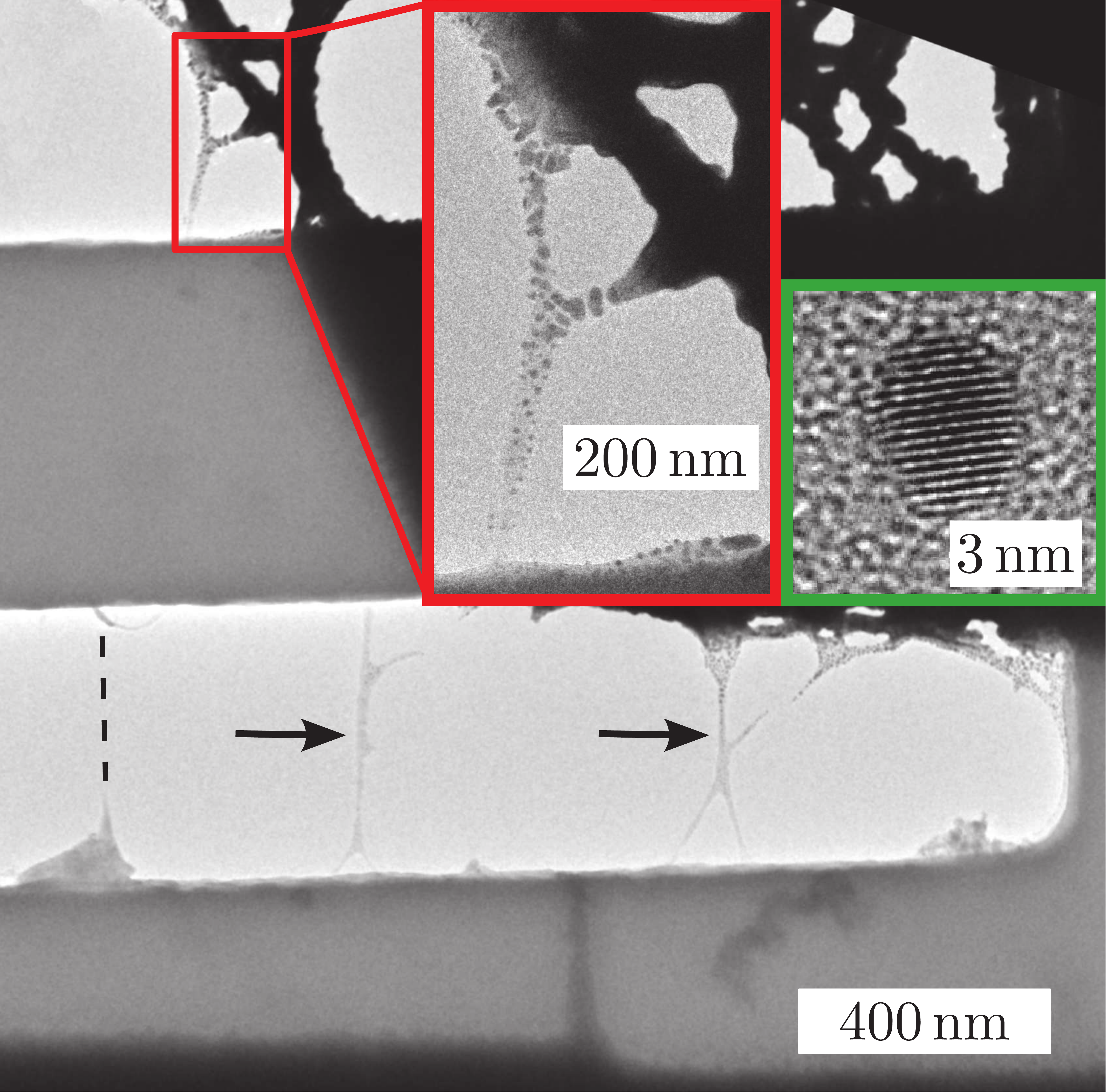}}\hfill
		\subfloat[Upper part: TEM image of the amorphous material bridging the slits. Lower part: TEM image of the device section marked in Fig.~\ref{REMpos30} with corresponding linescan averaging the entire image width.]{\label{TEMscan} 
			\includegraphics*[bb=0 0 410 410,width=.32\textwidth]{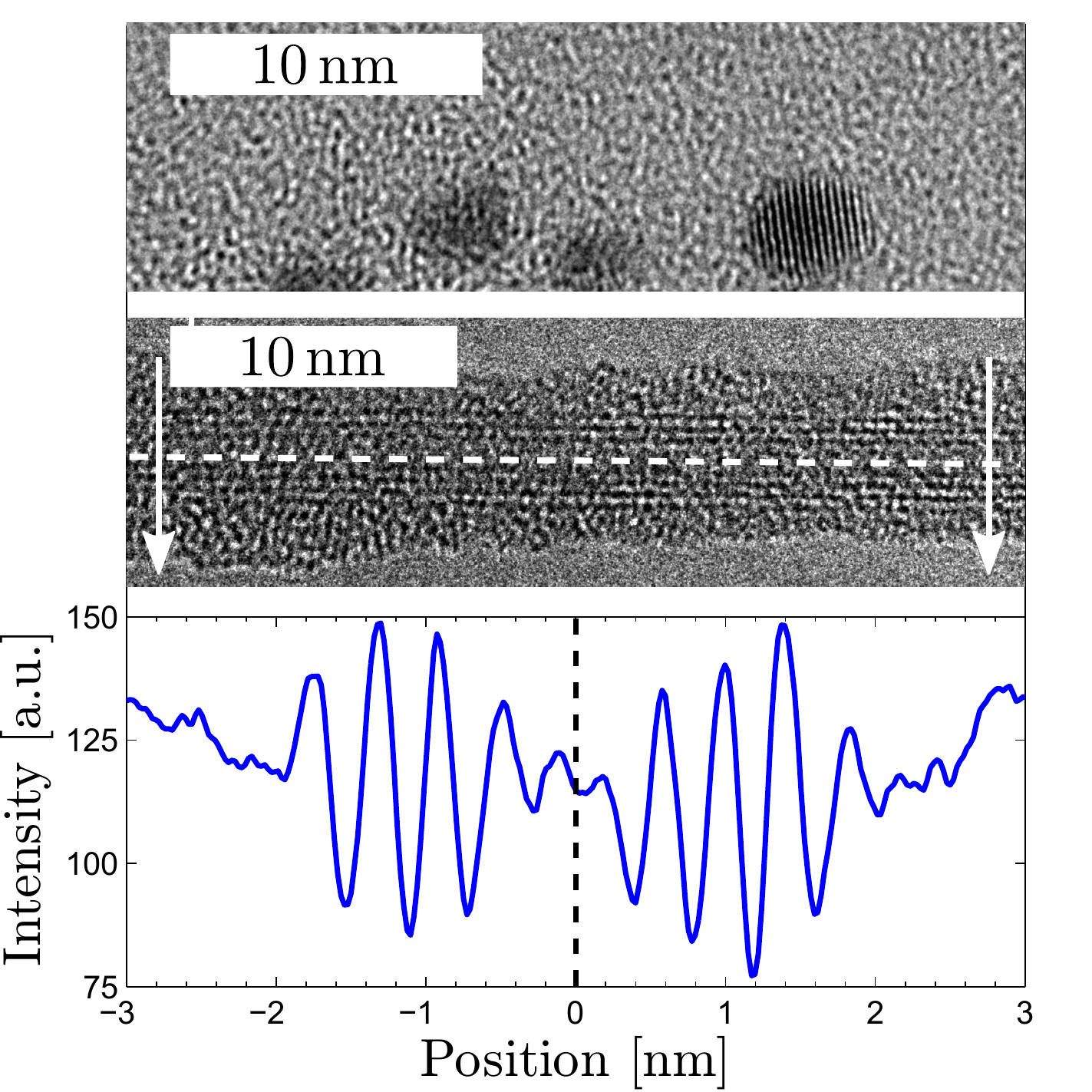}}\hfill
		\caption{SEM and HR-TEM micrographs of the contacted nanotube device.}
		\label{TEM}
	\end{figure*}	

	An SEM micrograph with larger magnification reveals that the CNT part across the bottom slit is covered with some material (see Fig.~\ref{REMpos30}). The upper contact reaches the nanotube in the middle of the top observation slit resulting in a diffuse interface. Additional structures that gradually fade out in the SEM image are bridging the bottom slit (see arrows in Fig.~\ref{REMpos30}). Furthermore, as can be seen in Fig.~\ref{Kontaktiert} compared to Fig.~\ref{CNTverlauf} the SEM contrast of the CNTs in contact with an electrode is strongly enhanced.
	
	The HR-TEM micrograph in Fig.~\ref{TEMgesamt} shows the same region as the SEM image in Fig.~\ref{REMpos30}. All structures bridging the slits appear with similar intensity. The structure containing the nanotube in the lower slit was destroyed by the TEM electron beam albeit its low acceleration voltage. At the interface of the contact to the CNT, many particles are embedded in an amorphous matrix of debris material (red inset). The particles have a crystalline structure as shown in the green inset taken from another sample region. Their lattice constant and contrast identifies them as gold droplets deposited onto the device during contact evaporation. 
	
	The upper HR-TEM micrograph in Fig.~\ref{TEMscan} is a closeup of a spiderweb-like structure bridging a slit similar to the ones marked with arrows in the SEM and TEM images in Figs.~\ref{REMpos30} and \ref{TEMgesamt}. It consists of amorphous material that hosts gold nanoparticles. The lower HR-TEM image is a zoom into the region marked with a blue rectangle in Fig.~\ref{REMpos30} taken before nanotube destruction. The same amorphous material as in the above image covers the nanotube. Nevertheless the CNT can still be observed as it is highlighted by the linescan in the bottom part of Fig.~\ref{TEMscan}. It is taken across the entire depicted nanotube section in the direction indicated with the white arrows. Three CNT shells can be revealed. Calculating the diameters by iteratively refining the shell diameters one obtains 1.19, 1.94, and 2.72\ph{nm} for the nanotube diameters. 
	
	\begin{figure}[htb]%
		\centering\includegraphics*[bb=0 0 362 252,width=.9\linewidth]{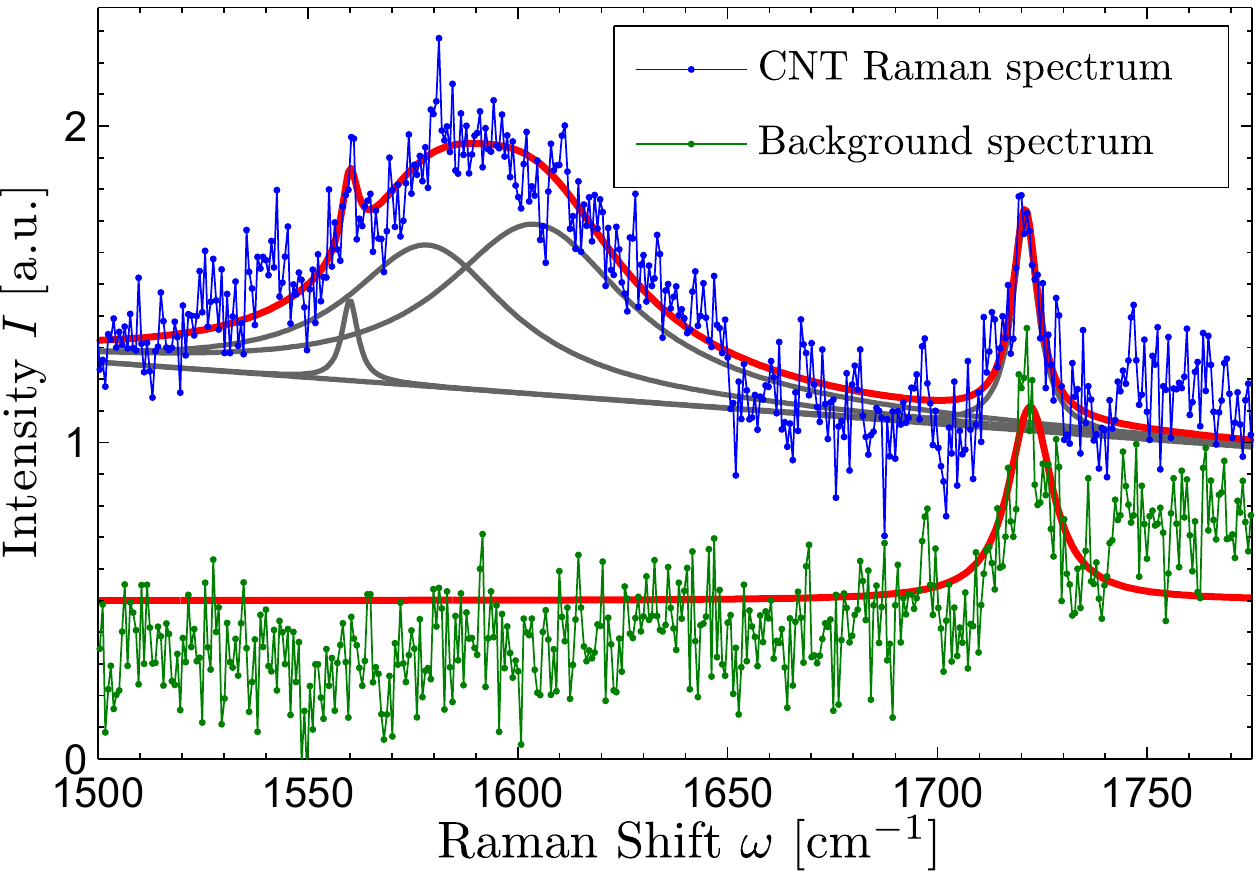}
		\caption{Raman measurement of the high-energy mode region. The Background spectrum is measured on a CNT-free part of the membrane.}
		\label{HEM}
	\end{figure}
	The high-frequency Raman spectrum of the device is presented in Fig.~\ref{HEM}. There are two distinct regions at high energies. First, there is the $G$-Mode of the CNT with three contributions: One $G^+$ band at $\omega_{G^+}=(1604\pm 1)\ph{cm^{-1}}$, a narrow $G^-$ band at $\omega_{G^-_S}=(1560\pm 1)\ph{cm^{-1}}$ with a width $\sigma_S=5\ph{cm^{-1}}$ and a broader one with $\omega_{G^-_M}=(1578\pm 1)\ph{cm^{-1}}$ and $\sigma_M=46\ph{cm^{-1}}$. 
	
	Of the three walls observed in TEM only two are in resonance with the laser light at $\lambda_0=488\ph{nm}$. The linewidths of the $G^-$ Raman modes of CNTs typically reflect the metallic or semiconducting character of a CNT with the broader one being metallic and the narrow one being semiconducting~\cite{Oron-Carl2005}. Following Piscanec \emph{et al.} \cite{Piscanec2007} we can assign diameters $d_S=(1.14\pm 0.05)\ph{nm}$ and $d_M=(2.6\pm 0.1)\ph{nm}$ for the semiconducting and metallic tube, respectively. 
	
	The second contribution to the Raman spectrum is a peak at $\omega_P=(1721\pm 1)\ph{cm^{-1}}$ that is also present in a membrane region without any CNTs. It can be attributed to the C=O bond in the ester carbonyl group of the PMMA electron beam resist. The absence of the unpolymerized C=C bond MMA peak at $\omega=1640\ph{cm^{-1}}$ (not shown) indicates a complete polymerization of the PMMA~\cite{Pallikari2001} which explains its resistance to standard removers as reported in section \ref{patterning}. We therefore believe that the material around the nanotube are PMMA remnants from an incomplete lift-off.

	\begin{figure}[htb]%
		\centering\includegraphics*[bb=0 0 338 246,width=.8\linewidth]{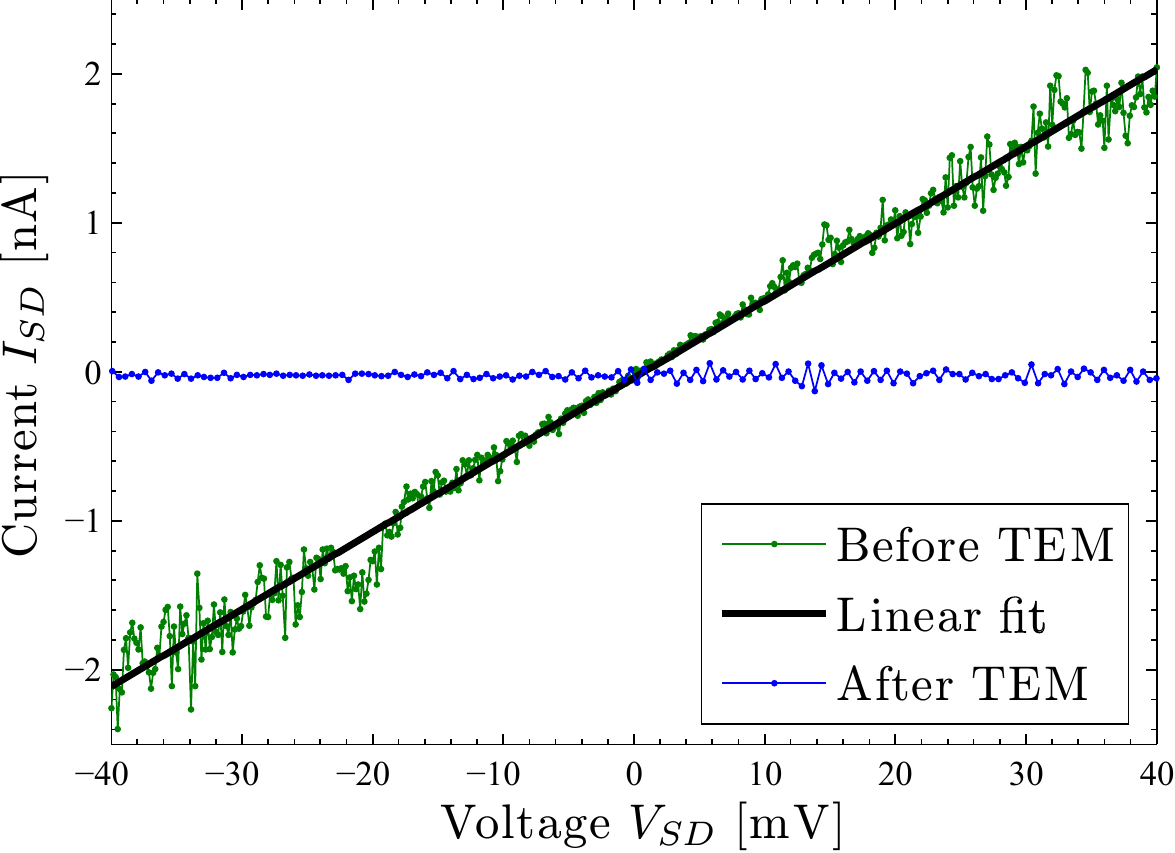}
		\caption{%
			Room-temperature current-voltage characteristics of the device plotted in Fig.~\ref{TEM}. It was destroyed during the TEM measurement.}
		\label{IV}
	\end{figure}
	The sample resistance is measured at room temperature in a probe station. Source and drain contacts are indicated in Fig.~\ref{Kontaktiert}. The CNT segment that is probed has a length of about 1.44\ph{\mu m}. It exhibits a linear current-voltage characteristics with a resistance $R\approx 19\ph{M\Omega}$ as plotted in Fig.~\ref{IV}. After destruction by the TEM electron beam the current between the electrodes is well below the detection limit of our instruments (cf. Fig.~\ref{IV}). This indicates that the current measured before was indeed carried by the nanotube device and that there are no parasitic resistances e.g. through the other structures bridging the lower slit.
	
	\section{Discussion}
	The ohmic behavior indicates a metallic chirality although the resistance value is about two orders of magnitude higher than expected~\cite{Nygard1999}. This feature can be explained by the results from other characterization methods. The diffusive upper contact interface is consisting mainly of the amorphous material covering the CNT (cf. Fig.~\ref{TEMscan}). This can lead to a high contact resistance. 
	
	The Raman measurements identify the material to be non-conductive PMMA. In the electron micrographs its insulating properties are revealed: Although the spider-web like structures are physically connected to the electrodes they are obscured by charging effects in the SEM as they are not electrically contacted. This does not happen in the TEM measurements. The strong contrast increase for contacted nanotubes between Figs.~\ref{CNTverlauf} and \ref{Kontaktiert} is a similar effect. Only CNTs with an electric contact can disperse the charges induced by the SEM into the electrodes. 
	
	Comparing the Raman and HR-TEM measurements we can conclude that the innermost tube is semiconducting while the outermost is metallic and therefore carries at least a part of the current. It is possible that the amorphous PMMA and the gold droplets which cover a wide range of the nanotube influence the transport through the outer shell. Unfortunately, we could not implement a four-terminal measurement due to a lack of space. Thus we cannot separate this effect from the contact resistance. 

	
\section{Summary and Conclusion}
	In this paper, we presented a comprehensive characterization of a CNT transport device by combining electron microscopy and Raman spectroscopy with electronic transport measurements. To obtain these complementary information about the investigated device we introduced a sample design that is based upon standard electron beam lithography on a commercial TEM membrane of 200\ph{nm} thick Si$_3$N$_4$. 
	
	Connecting the results obtained by the different methods we find that despite the low growth temperature the CNT consists of three walls. The innermost shell is semiconducting while the outermost is metallic. This is consistent with the ohmic transport behavior. The high resistivity can be explained with a large contact resistance due to coverage with PMMA leading to an undefined contact interface.
	
	Besides the determination of the morphology of a CNT device it will be possible to unambiguously quantify the degree of its functionalization using this technique. Fullerene filling, leading to so-called peapods, can be imaged as well as single (magnetic) molecules sticking on the outside wall of the CNTs. It also extends to other material systems as nanowires where it could help to determine the influence of stacking faults on the electronic transport. In a next step we will try to correlate the electronic structure of an individual CNT quantum dot obtained by low-temperature transport measurements with its atomic structure measured in TEM.

\begin{acknowledgement}
	We thank Ren\'e Borowski for RIBE processing and the DFG (Forschergruppe FOR912) for financial support.
\end{acknowledgement}

%
\bibliographystyle{pss}

\providecommand{\WileyBibTextsc}{}
\let\textsc\WileyBibTextsc
\providecommand{\othercit}{}
\providecommand{\jr}[1]{#1}
\providecommand{\etal}{~et~al.}


\begin{thebibliography}{[10]}

\bibitem{Spudat2009}
 \textsc{C.~Spudat},  \textsc{C.~Meyer},  \textsc{K.~Goss},  and
  \textsc{C.\,M. Schneider},
 \jr{physica status solidi (b)} \textbf{246}(11-12), 2498--2501 (2009).


\bibitem{Qi2003}
 \textsc{P.~Qi},  \textsc{O.~Vermesh},  \textsc{M.~Grecu},  \textsc{A.~Javey},
  \textsc{Q.~Wang},  \textsc{H.~Dai},  \textsc{S.~Peng},  and  \textsc{K.\,J.
  Cho},
 \jr{Nano Letters} \textbf{3}(3), 347--351 (2003).


\bibitem{Spudat2010}
 \textsc{C.~Spudat},  \textsc{M.~M\"uller},  \textsc{L.~Houben},
  \textsc{J.~Maultzsch},  \textsc{K.~Goss},  \textsc{C.~Thomsen},
  \textsc{C.\,M. Schneider},  and  \textsc{C.~Meyer},
 \jr{Nano Lett} \textbf{10}(11), 4470--4474 (2010).


\bibitem{Eliasen2010}
 \textsc{A.~Eliasen},  \textsc{J.~Paaske},  \textsc{K.~Flensberg},
  \textsc{S.~Smerat},  \textsc{M.~Leijnse},  \textsc{M.\,R. Wegewijs},
  \textsc{H.\,I. J\o{}rgensen},  \textsc{M.~Monthioux},  and
  \textsc{J.~Nyg\aa{}rd},
 \jr{Phys. Rev. B} \textbf{81}(15), 155431 (2010).


\bibitem{Goss2010}
 \textsc{K.~{Go{\ss}}},  \textsc{S.~{Smerat}},  \textsc{M.~{Leijnse}},
  \textsc{M.\,R. {Wegewijs}},  \textsc{C.\,M. {Schneider}},  and
  \textsc{C.~{Meyer}},
 \jr{arXiv:1011.4004v1} (2010).


\bibitem{Langer1996}
 \textsc{L.~Langer},  \textsc{V.~Bayot},  \textsc{E.~Grivei},  \textsc{J.\,P.
  Issi},  \textsc{J.\,P. Heremans},  \textsc{C.\,H. Olk},
  \textsc{L.~Stockman},  \textsc{C.~Van~Haesendonck},  and
  \textsc{Y.~Bruynseraede},
 \jr{Phys. Rev. Lett.} \textbf{76}(3), 479--482 (1996).


\bibitem{Haider1998}
 \textsc{M.~Haider},  \textsc{S.~Uhlemann},  \textsc{E.~Schwan},
  \textsc{H.~Rose},  \textsc{B.~Kabius},  and  \textsc{K.~Urban},
 \jr{Nature} \textbf{392}(6678), 768--769 (1998).


\bibitem{Meyer2009}
 \textsc{C.~Meyer},  \textsc{C.~Spudat},  \textsc{L.~Houben},  and
  \textsc{C.\,M. Schneider},
 \jr{Nanotechnology} \textbf{20}(6), 065603 (2009).


\bibitem{Oron-Carl2005}
 \textsc{M.~Oron-Carl},  \textsc{F.~Hennrich},  \textsc{M.\,M. Kappes},
  \textsc{H.\,v. L\"ohneysen},  and  \textsc{R.~Krupke},
 \jr{Nano Letters} \textbf{5}(9), 1761--1767 (2005).


\bibitem{Piscanec2007}
 \textsc{S.~Piscanec},  \textsc{M.~Lazzeri},  \textsc{J.~Robertson},
  \textsc{A.\,C. Ferrari},  and  \textsc{F.~Mauri},
 \jr{Phys. Rev. B} \textbf{75}(3), 035427 (2007).


\bibitem{Pallikari2001}
 \textsc{F.~Pallikari},  \textsc{G.~Chondrokoukis},  \textsc{M.~Rebelakis},
  and  \textsc{Y.~Kotsalas},
 \jr{Materials Research Innovations} \textbf{4}, 89--92 (2001).


\bibitem{Nygard1999}
 \textsc{J.~{Nyg\aa{}rd}},  \textsc{D.\,H. {Cobden}},  \textsc{M.~{Bockrath}},
  \textsc{P.\,L. {McEuen}},  and  \textsc{P.\,E. {Lindelof}},
 \jr{Applied Physics A: Materials Science \& Processing} \textbf{69}, 297--304
  (1999).


\end{thebibliography}
%
%
\providecommand{\WileyBibTextsc}{}
\let\textsc\WileyBibTextsc
\providecommand{\othercit}{}
\providecommand{\jr}[1]{#1}
\providecommand{\etal}{~et~al.}

\end{document}